\documentclass[referee]{raa}           

\usepackage{graphicx,times}
\usepackage{natbib}
\usepackage{amssymb,amsmath}
\bibpunct{(}{)}{;}{a}{}{,}

\usepackage[pagebackref=true]{hyperref}

\begin{document}

   \title{Spectral fittings of warm coronal radiation with high seed photon temperature: apparent low-temperature and flat soft excess in AGNs}

 \volnopage{ {\bf 20XX} Vol.\ {\bf X} No. {\bf XX}, 000--000}
   \setcounter{page}{1}

   \author{Ze-Yuan Tang \and Jun-Jie Feng \and Jun-Hui Fan \inst{}
   }

\institute{Center for Astrophysics, Guangzhou University, Guangzhou 510006, People’s Republic of China {\it tangzy@mail.ustc.edu.cn}\\
        \and
             Astronomy Science and Technology Research Laboratory of Department of Education of Guangdong Province, Guangzhou 510006, People’s Republic of China\\
        \and
             Key Laboratory for Astronomical Observation and Technology of Guangzhou, Guangzhou 510006, People’s Republic of China\\
\vs\no
   {\small Received 20xx month day; accepted 20xx month day}}

\abstract{A warm corona has been widely proposed to explain the soft X-ray excess (SE) above the 2--10 keV power law extrapolation in AGNs. In actual spectral fittings, the warm coronal seed photon temperature ($T_{\rm s}$) is usually assumed to be far away from the soft X-ray, but $kT_{\rm s}$ can reach close to 0.1 keV in standard accretion disc model. In this study, we used Monte Carlo simulations to obtain radiation spectra from a slab-like warm corona and fitted the spectra using the spherical-geometry-based routine \textsc{thcomp} or a thermal component. Our findings reveal that high $T_{\rm s}$ can influence the fitting results. A moderately high $kT_{\rm s}$ (around 0.03 keV) can result in an apparent low-temperature and flat SE, while an extremely high $kT_{\rm s}$ (around 0.07 keV) can even produce an unobserved blackbody-like SE. Our conclusions indicate that, for spectral fittings of the warm coronal radiation (SE in AGNs), $kT_{\rm s}$ should be treated as a free parameter with an upper limit, and an accurate coronal geometry is necessary when $kT_{\rm s}>0.01$ keV.
\keywords{ black hole physics --- galaxies: active --- accretion, accretion discs
}
}

   \authorrunning{Z.-Y. Tang et al. }            
   \titlerunning{High $T_{\rm s}$ warm coronal spectra}  
   \maketitle

%
\section{Introduction}
In soft X-ray band ($\sim0.3-2$ keV), an excess (soft excess, SE) above the 2-10 keV power law extrapolation is found in the majority of Active Galactic Nuclei (AGNs) spectra \citep{Walter1993,Boissay2016,Liu2017}. There are two prominent explanations for the origin of the soft excess (SE). One possibility is that the SE arises from the ionized reflection of the disc's hot coronal flux \citep{Ross2005,Crummy2006}. However, this ionized reflection model necessitates unusually high black hole spin, compact hot coronae, and sometimes high disc densities to account for relativistic effects and smudge out line emissions \citep{Crummy2006,Jiang2018,Garcia2019,XuYR2021}. Furthermore, numerous observational analyses indicate a weak correlation between the SE and reflection ratio \citep{Mehdipour2011,Mehdipour2015,Noda2013,Matt2014,Boissay2016,Porquet2018,Matzeu2020,Xu2021}. As a result, these works preferred an alternative interpretation which consistently provides an excellent fit. According to this interpretation, the SE formation is attributed to the Comptonisation of extreme UV (EUV) disc photons in an optically-thick (with $\tau \sim$ 10–40) and warm (with $kT \sim$ 0.1–1 keV) plasma, i.e., ``warm corona" \citep{Done2012,Petrucci2013}. According to studies on variability \citep{Marco2013,Kara2016,Mallick2021,Zoghbi2023} and theory \citep{Ballantyne2020,Gronkiewicz2023}, the scale of the warm corona can range from a few to a few tens of gravitational radii ($r_{\rm g}$).

In warm coronal spectral fittings, the SE is usually fitted by the Comptonisation spectra produced by the routine \textsc{nthcomp} \citep{Zdziarski1996,Zycki1999}, or the new version \textsc{thcomp} \citep{Zdziarski2020}. However, the EUV is very heavily extinguished by the interstellar medium of our Galaxy, and therefore only the high energy part ($\gtrsim 0.3$keV, i.e., the SE) of the Comptonisation spectra is visible. As a result, the spectral fittings can constrain the warm coronal parameters at most, leaving the seed photon temperature ($T_{\rm s}$) uncertain.

Although $T_{\rm s}$ is weakly known, we still need a specific value in spectral fittings. In actual works, the easiest way is to set $kT_{\rm s}$ much lower than the magnitude of 0.1 keV \citep[e.g. $kT_{\rm s}$=3 eV,][]{Petrucci2018}, then the SE profile will be weakly dependent on $T_{\rm s}$ as photons in the visible part ($\gtrsim 0.3$keV) have been multiple-scattered and lose their initial information. A very low $kT_{\rm s}$ is convenient for studying the warm coronal structure, but it would require an unphysically large warm coronal scale to maintain self-consistency \citep{Petrucci2013, Petrucci2018}. A more physical idea is that at each radius ($r$) the $T_{\rm s}(r)$ follows the local disc temperature $T_{\rm d}(r)$, which could be close to the magnitude of 0.1 keV \citep{Shakura1973,li2010,Reynolds2020}. \cite{Done2012} suggested that the disc can be radially divided into two parts: in the outer part, the radiation is semi-blackbody-like, i.e., blackbody radiation ($B_{\nu}(T)$) with a colour temperature correction ($f_{\rm col}$): $I_{\nu}=B_{\nu}(f_{\rm col}T_{\rm d})/f_{\rm col}^4~$; in the inner part, the radiation has been Comptonised by the warm and hot coronae, and the $T_{\rm s}(r)$ is the same as the $T_{\rm d}(r)$ predicted by the standard accretion disc model \citep{Shakura1973}. The corresponding public code \textsc{optxagnf} generates the Comptonisation component by \textsc{nthcomp}, and followed by some other works \citep[e.g.][]{Kubota2018,Kubota2019,Hagen2023}. However, recently many simulation works suggested that the warm corona should have a vertical structure deep inside \citep{Rozanska2015,Petrucci2020,Ballantyne2020,Ballantyne2020b,Gronkiewicz2023},and therefore an $f_{\rm col}$ should also be considered in the Comptonisation region and perhaps the corresponding $T_{\rm s}$ is underestimated.

Modeling astrophysical Comptonisation spectra faces challenges at high seed photon temperatures due to limitations of traditional theories and numerical models. The Kompaneets equation, which underlies codes like \textsc{nthcomp} and \textsc{thcomp}, is strictly valid under two conditions: (1) $h\nu \ll kT_{\rm e}$ and (2) $kT_{\rm e} \ll m_{\rm e}c^2$ \citep{Kompaneets1957, Sunyaev1980}. While \textsc{nthcomp/thcomp} have significantly improved the calculation accuracy when $kT_e\sim m_e c^2$ by including the Klein-Nishina scattering cross section, it retains some inaccuracies in calculation when $h\nu$ approaches $kT_{\rm e}$ \citep[][due to the expression of Klein-Nishina formula, see Eq.~(\ref{eq:KN})]{Zdziarski1996,Zycki1999,Zdziarski2020}. The calculation errors will accumulate when $T_{\rm s}$ approaches $T_{\rm e}$, as there are substantial scatterings at $h\nu \sim kT_{\rm e}$. Furthermore, \textsc{nthcomp/thcomp} adopt a spherical geometry with a sinusoidal seed photon distribution ($\propto(\tau_0/\pi\tau)\sin (\pi\tau/\tau_0)$, where $\tau_0$ is the total radial optical depth) for computational expedience \citep[they followed][]{Sunyaev1980}, but the actual warm corona is perhaps a slab-like plasma with photons from the bottom \citep{Rozanska2015, Petrucci2018, Petrucci2020, Ballantyne2020, Ballantyne2020b, Gronkiewicz2023}. These differences will further magnify the existing modeling errors. As a result, at high seed photon temperature $T_s$ approaching $T_e$, the fitted warm coronal temperatures and spectral indices are not reliable.

Several recent studies of slab-like warm coronae have involved (effective) high-temperature seed photons \citep[e.g.][]{Petrucci2020, Gronkiewicz2023}. However, these studies adopt a vertically temperature-varying plasma which makes defining a representative global temperature difficult. The varying temperature means the final results can not be straightforwardly estimated, as the final apparent warm coronal temperature depend on the scattering order and can deviate from the optical-depth-weighted average temperature. It is therefore challenging to isolate the influence of high seed photon temperatures on the fitted warm coronal parameters when using a vertically temperature-varying model. Adopting a uniform temperature slab allows the true temperature to be known, which enables us to identify the effects caused by seed photon temperature. In observational analyses, when using codes like NTHCOMP/THCOMP, the constant warm coronal temperature is already assumed by default \citep[e.g.][]{Jin2012,Porquet2018,Matzeu2020,Xu2021,Jin2022}. The fitted temperature then provides a reference for the average warm coronal temperature.

In this study, we utilize the public Monte Carlo simulation code \textsc{grmonty} \citep{Dolence2009} to obtain radiation spectra for the warm coronal region. Subsequently, we fit these spectra using either \textsc{thcomp} or a thermal component. Then, by assuming a uniform scattering slab in simulations, we have a priori knowledge of the temperature, allowing us to determine the effectiveness of the fitting methods.

The paper is organized as follows: The simulation and fitting methods are introduced in Sec.~\ref{sec:method}. In Sec.~\ref{sec:num}, we show the numerical results and study how the high seed photon temperature affects the SE profile. In Sec.~\ref{sec:discuss}, we discuss the possible effects on actual spectral fittings. The conclusions are show in Sec.~\ref{sec:conclu}.

\section{Simulation and fitting methods}
\label{sec:method}
\subsection{Simulation methods}
The physical scenario we simulated is straightforward: photons freely traverse within a slab-like electron gas until they undergo scattering; this process is repeated until the photons either escape or are absorbed. Our Monte Carlo simulation kernel for Compton scattering is from the public code \textsc{grmonty} \citep{Dolence2009}. The kernel uses the Klein-Nishina differential cross section:
\begin{equation}
	\label{eq:KN}
	\frac{2\pi}{\sigma_{\rm T}}\frac{\mathrm{d}\sigma_{\rm K N}}{\mathrm{d}\epsilon^{\prime}}=\frac{1}{\epsilon^{2}}(\frac{\epsilon}{\epsilon^{\prime}}+\frac{\epsilon^{\prime}}{\epsilon}-\sin^{2}\theta),
\end{equation}
where $\sigma_{\rm T}$ is the Thomson cross section, $\theta$ is the scattering angle, and $\epsilon$, $\epsilon^{\prime}$ are the initial, final photon energy in the electron rest frame, respectively. It is worth noting that when $\epsilon\sim\epsilon^{\prime}$, the three terms in right-hand side of Eq.~\ref{eq:KN} have comparable contributions to the cross section, makes semi-analytical approximations very difficult at this time, and therefore we need Monte Carlo simulations. 

The configuration of the disc-corona system resembles a slab. The disc occupies a semi-infinite region at the bottom, while the warm corona forms a layer on top. The disc has low temperatures and high densities, and therefore absorption dominates; whereas the warm corona has high temperatures and low densities, and therefore scattering dominates \citep{Done2012, Petrucci2018, Petrucci2020, Ballantyne2020, Gronkiewicz2023}. Seed photons are emitted isotropically from the disc and undergo scattering processes within the warm coronal region. We assume that the warm corona is a grey atmosphere with pure scattering, and the electrons within are in a thermal distribution. The photons will escape from the system when they cross the upper boundary (the warm coronal surface). Some photons will return to the disc, most of them will be absorbed but a few of them will be scattered back to the warm corona again. The scattered-back photons can be effectively considered as a special form of emitted photons, and then the overall effect caused by the absorption-scattering competition can be considered as an increase in the seed photon temperature \citep[e.g.][]{Done2012}. In our simulations, a photon will be be absorbed immediately after returning to the disc, it means that our setting $T_{\rm s}$ is indeed an effective temperature.

\begin{figure*}
	\centering
	\includegraphics[width=1.0\textwidth]{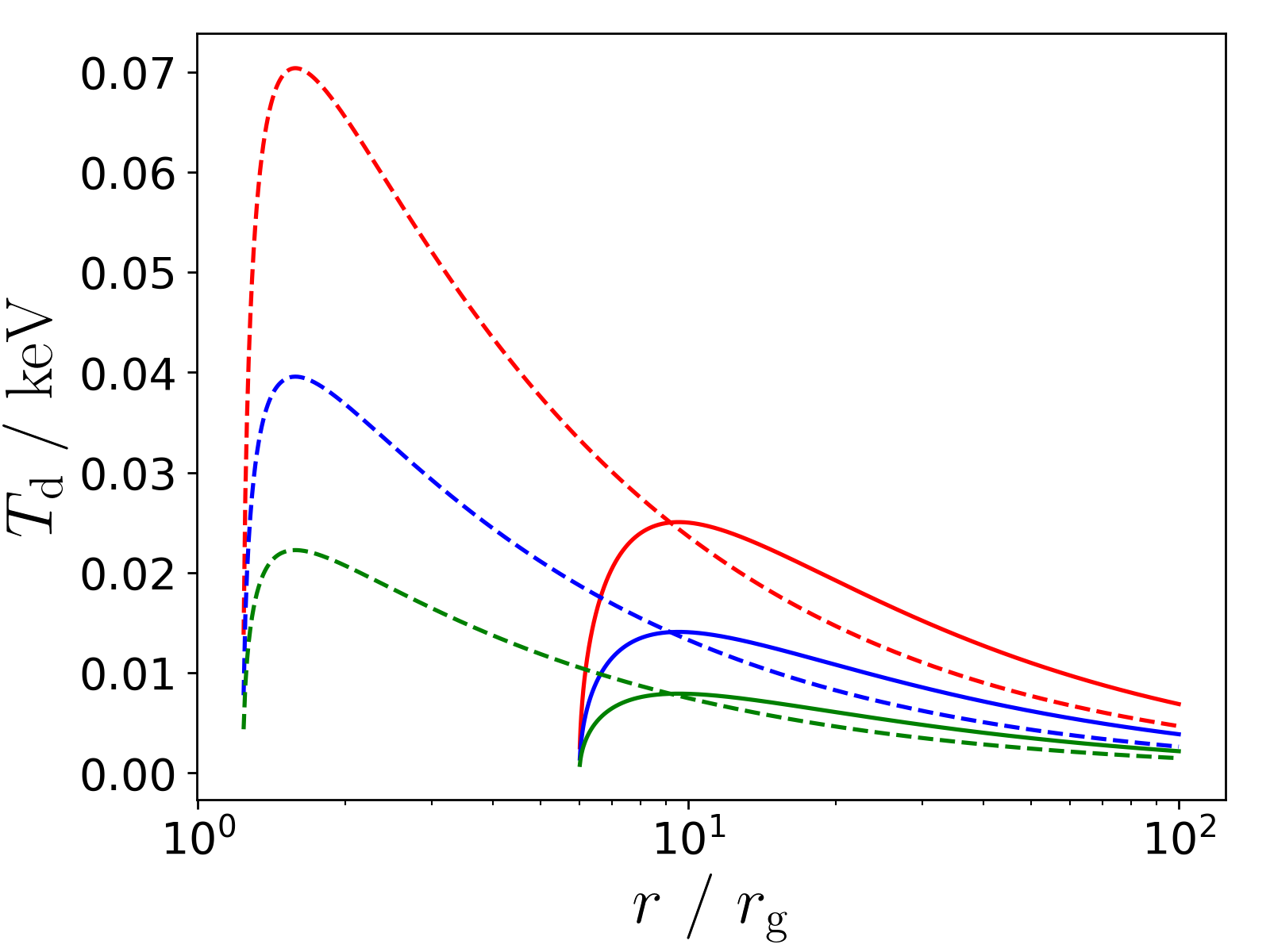}
	\caption{Temperature profiles of relativistic standard accretion discs. The disc temperature $T$ is in unit of keV, and the radius $r$ is in unit of gravitational radius $r_{\rm g}$. The black hole mass ($M_{\rm BH}$) is $10^6$ (red), $10^7$ (blue) or $10^8$ (green), in unit of $M_{\odot}$. The black hole spin is 0 (solid) or 0.998 (dashed). The luminosity ($L$) is 0.3 Eddington luminosity. Readers can scale out the profiles by the relation $T\propto(L/M_{\rm BH})^{1/4}$.} 
	\label{fig: disct}
\end{figure*}

Now we introduce the method for selecting the input seed photon temperature. Due to the uncertainties in the disc-corona structure, simulation and theoretical works prefer a single temperature as input \citep[e.g.][]{Petrucci2013, Petrucci2020, Ballantyne2020, Gronkiewicz2023}. The final spectrum is a superposition of many spectra with different $T_{\rm s}$. Its ultimate characteristics would approach the qualitative properties of the spectrum with a typical $T_{\rm s}$ within the dominant range. Meanwhile, it is believed that $T_{\rm s}$ would not significantly deviate from $T_{\rm d}$, and therefore the simulation and theoretical works still consider the original $T_{\rm d}$ as an indication of $T_{\rm s}$. The peak temperatures of AGN discs range from 0.01 to 0.1 keV \citep[e.g.][]{Reynolds2020}, while the outer disk temperatures can be as low as a few eVs or even less. To give readers an intuitive impression, a few typical temperature profiles for a relativistic standard accretion disc \citep{Novikov1973} are shown in Fig.~\ref{fig: disct}, readers can also scale out the profiles by the relation $T\propto(L/M_{\rm BH})^{1/4}$ (where $L$ is the disc luminosity and $M_{\rm BH}$ is the black hole mass). Due to time limitations for simulations, we aim to use as few input values as possible to obtain spectra that represent the main characteristics across different energy bands and cover the whole energy range. We conducted some preliminary fast simulations to assist with $T_{\rm s}$ selection. Firstly, for all $kT_{\rm s}\lesssim0.01$ keV, the SE ($\gtrsim0.3$ keV) profile is not affected by $T_{\rm s}$, and therefore we set the lowest $kT_{\rm s}$ in the simulations to 0.01 keV. Secondly, for $kT_{\rm s}\gtrsim0.07$ keV, considerable SE profiles will be blackbody-like, and therefore we set the highest $kT_{\rm s}$ in the simulations to 0.07 keV. Finally, the current two $kT_{\rm s}$ values are far apart, and therefore we choose an intermediate value of 0.03 keV, at which the SE profile is affected by $T_{\rm s}$ but not blackbody-like. 0.01 keV is a typical peak temperature of ultra-high-mass and low/moderate-accretion-rate AGNs \citep[e.g. early quasars,][]{Wu2015}, or $T_{\rm d}$ at larger radii of AGNs with higher peak temperature; 0.07 keV corresponds to the peak temperature of low-mass and high-accretion-rate AGNs \citep[e.g. Narrow-line Seyfert 1 galaxies, NLS1s,][]{Osterbrock1985, Gu2015}; 0.03 keV corresponds to a variety of situations, such as $T_{\rm d}$ at $\sim 10r_{\rm g}$ of NLS1s, peak temperature of some luminous quasars \citep[high-mass and high-accretion-rate, e.g.][]{Laurenti2022}, etc.

Our main concern is the warm coronae with low-temperature electrons ($kT_{\rm e}\in[0.1, 0.4]~{\rm keV}$). Meanwhile, the warm coronal vertical Thomson optical depth $\tau\in[5, 70]$. Observations and simulations show that these selected value ranges cover most of the corona \cite[see][and references therein]{Gronkiewicz2023}. The warm coronae outside of these ranges will not affect our results, because they are few in number and their parameter values are always on the same side of the median.

\subsection{Fitting methods}
To emulate the actual detectors, the output spectra are divided into frequency channels ($I$) of logarithmic width $\Delta \log \nu=0.082$, and the detectors only tell us the photon counts in each channel ($C(I)$). The X-ray detectors can not receive the distant photons with energies below $\sim 0.3$ keV because of interstellar absorption. Considering the specific frequencies of the channels in simulations, we set the lower boundary of the fitting band as 0.26 keV, the value only weakly influences our results as long as the high energy cut-off of SE is visible or not obvious. Meanwhile, in studies of the warm corona, one approach is to analyze the properties of the remaining soft X-ray flux (i.e., SE) after removing the background power-law component. Considering the signal-to-noise ratio, in actual observational works, the upper energy limit of SE is usually set between $\sim$1 keV \citep[e.g.][]{Boissay2016,Zoghbi2023} to $\sim$2 keV \citep[e.g.][]{Laurenti2022,Tang2023}, {varying in different cases. Therefore, we set the upper boundary of the fitting band as $1.0,~1.2,~2.1$ keV in our fittings.

We use two different methods to fit the output spectra. In the first method we use a thermal component to fit the spectra and get best-fitting blackbody temperatures ($T_{\rm b}$). In the second method we make use of the warm corona model. As \textsc{thcomp} requires, the warm coronal structure is described by warm coronal temperature $T_{\rm wc}$ and photon index $\Gamma$ ($\Gamma$ is a function of $T_{\rm wc}$ and $\tau_{\rm wc}$). We further set the input photons in a thermal distribution of $T_{\rm d}$. All the seed photons will be Comptonised in the warm coronal rest frame. Then, \textsc{thcomp} has the capability to generate Comptonisation spectra by employing a sinusoidal distribution of seed photons within spherical electron gas. We set $kT_{\rm d}=0.01,~0.03,~0.07$ keV in fittings, and since we do not have a priori disc temperature in observational works, $T_{\rm d}$ does not need to match the seed photon temperature $T_{\rm s}$. Following the other warm coronal works \citep[e.g.][]{Waddell2023}, we set $kT_{\rm wc}\in[0.1,~1]$ keV and $\Gamma\in[2.0,~3.5]$.

Following \textsc{xspec} \citep{XSPEC}, the fit statistic in use for determining the best-fitting model is $\chi^{2}$:
\begin{equation}
	\chi^{2}=\sum\frac{(C(I)-C_{\rm m}(I))^{2}}{(\sigma(I))^{2}},
\end{equation}
where $C_{\rm m}$ is the model-predicted photon count and $\sigma(I)$ is the simulation error. $\sigma(I)$ is estimated by $\sqrt{C(I)}$ in this work.

We further define a ratio:
\begin{equation}
	R_{\rm b/c}=\frac{\chi^{2}_{\rm b}}{\chi^{2}_{\rm wc}},
\end{equation}
where the subscripts ${\rm b,~wc}$ correspond to the blackbody, warm corona fitting methods, respectively.

\section{Numerical results}
\label{sec:num}
\subsection{Output spectra}

We first visualize the output spectra and qualitatively analyze the results. Since the photon distribution in a channel is unknown, we approximate the distribution to be uniform and set the energy of the plot point at the geometric mean of the channel boundaries: $E_{\rm p}=\sqrt{E_{\rm min}(I)E_{\rm max}(I)}$. 

\begin{figure*}
	\centering
	\includegraphics[width=1.0\textwidth]{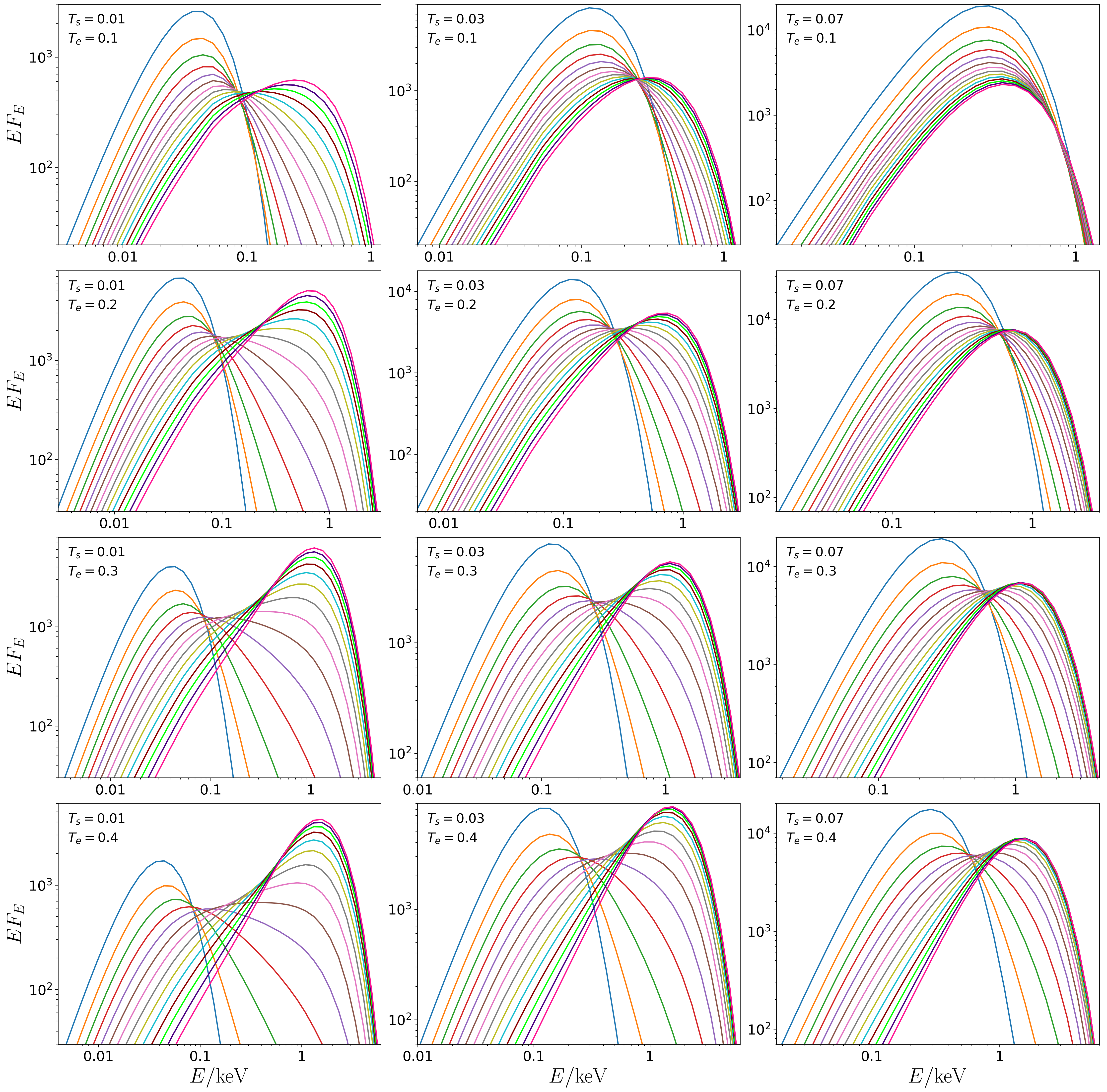}
	\caption{The output spectra of numerical simulations. $E$ of the plot point for a channel is at the geometric mean of the channel boundaries: $E_{\rm p}=\sqrt{E_{\rm min}(I)E_{\rm max}(I)}$. From left to right columns: seed photon temperature $kT_{\rm s}=0.01,~0.03,~0.07~{\rm keV}$. From top to bottom rows: electron temperature $kT_{\rm e}=0.1,~0.2,~0.3,~0.4~{\rm keV}$. In each panel, the curves from left to right correspond to the vertical Thomson optical depth $\tau=5,~10,~15,\cdot\cdot\cdot,70$, respectively. From bottom-left to top-right panels, the spectra will transition from \textsc{thcomp}-like spectra to blackbody-like spectra.} 
	\label{fig: spec}
\end{figure*}

The output spectra are shown in Fig.~\ref{fig: spec}\footnote{In this paper, we ignore the symbol $k$ of Boltzmann constant in all the figures for simple layout.}. There are only three parameters: $T_{\rm s}$ (decides the low energy cut-off), $T_{\rm e}$ (decides the high energy cut-off) and $\tau$. In general, when the two energy cut-offs are further from each other (i.e., $T_{\rm e}/T_{\rm s}$ is larger), the spectra are more likely similar to the ones produced by \textsc{thcomp}. For example, in the left-bottom panel ($kT_{\rm s}=0.01,~kT_{\rm e}=0.4$), for the spectra corresponding to moderate optical depths (middle curves), they show power-law profiles between their low and high energy cut-offs; while for the spectra corresponding to very small or large optical depth (left or right curves respectively), the power-law profiles are so steep that the energy rollovers are not very obvious, but the power-law part can still be distinguished by careful observation. In contrast, none of the spectra in the right-top panel ($kT_{\rm s}=0.07,~kT_{\rm e}=0.1$) contains a clear power-law profile, and in fact they look like a thermal bump. It seems that when the two energy cut-offs are close to each other, there will be only a narrow frequency window left to form a power-law component. As a result, when $T_{\rm e}/T_{\rm s}$ drops (from left-bottom to right-top panels), the window widths reduce and the spectra are more and more blackbody-like.

In every panel of Fig.~\ref{fig: spec}, one can see that the evolution trend of the spectral profile with the optical depth is continuous and monotonic, and the trends are very similar in different panels. As shown, when $\tau$ increases, the power-law photon index $\Gamma$ (if there exists) decreases, and the energies of the two cut-offs increase. Meanwhile, the low energy cut-off increases faster than the high one, which makes all the spectra corresponding to high $\tau$ more or less look like a thermal bump. One main difference between the trends should be mentioned here: for high $T_{\rm e}$, the scattering is more efficient, and therefore the evolution of the photon index and energy cut-offs will be more dramatically. As a result, the spectra with a small $T_{\rm e}/T_{\rm s}$ may not turn to blackbody-like at a small $\tau$.

\subsection{Spectral fittings}

\begin{figure*}
	\centering
	\includegraphics[width=1.0\textwidth]{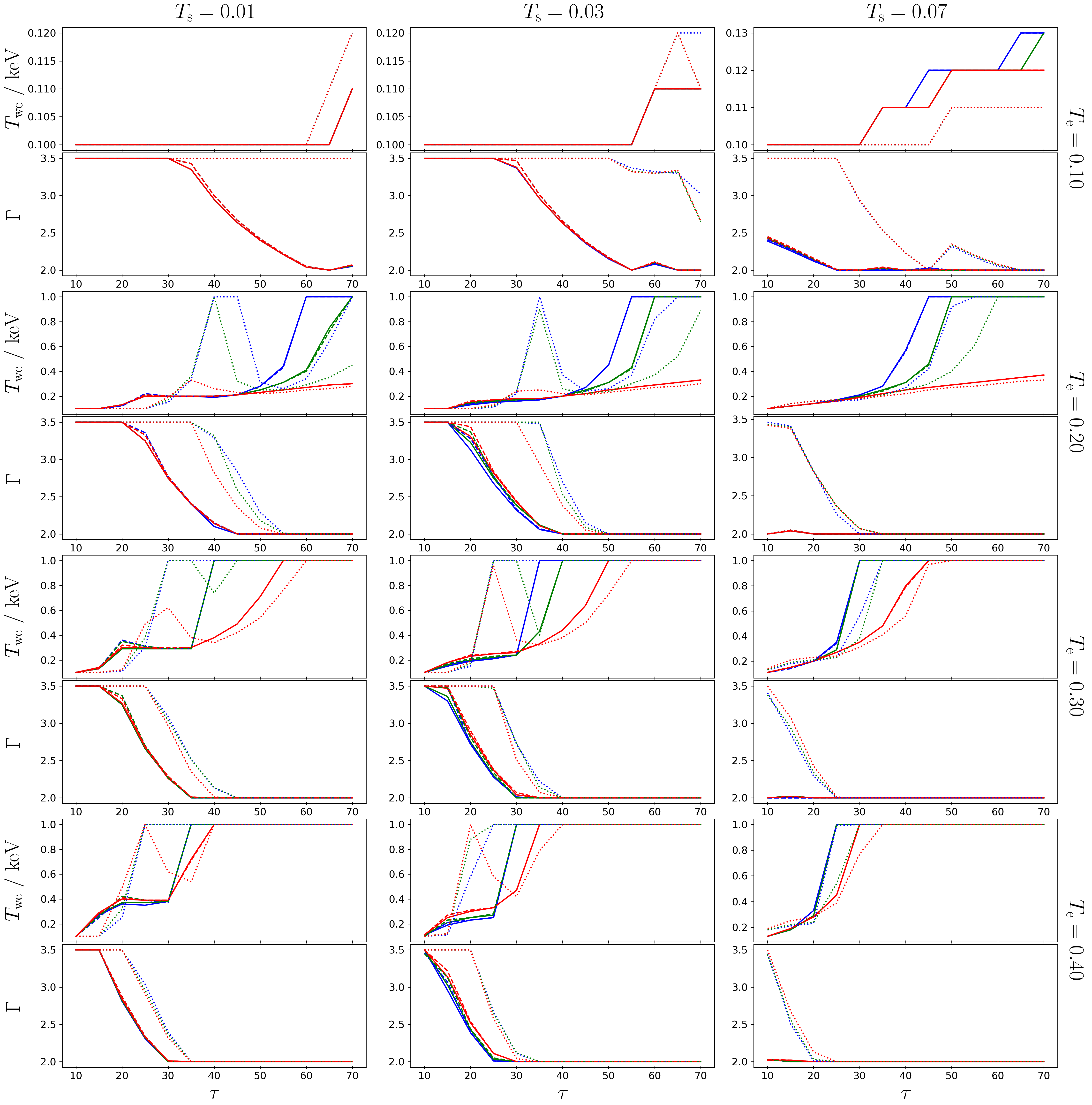}
	\caption{Best-fitting spectral parameters of the numerical simulation results, measured by assuming a warm corona model. $\tau$ is the vertical Thomson optical depth. The results of one simulation case are shown in two panels ($1\times2$, $T_{\rm wc}$ and $\Gamma$). The simulation parameters are $kT_{\rm s}=0.01,~0.03,~0.07$ keV (from left to right cases), and $kT_{\rm e}=0.1,~0.2,~0.3,~0.4$ keV (from top to bottom cases). The fitting disc temperature $kT_{\rm d}=$ 0.01 (solid), 0.03 (dashed), 0.07 (dotted) keV, and the upper boundaries of the fitting band are 1.0 (blue), 1.2 (green), 2.1 (red) keV. Noticed that the curve number (9) is the same in each panel, though part of the curves are not visible due to high overlap in some panels. In the reasonable range, $T_{\rm wc}$ matches $T_{\rm e}$ when $kT_{\rm s}=0.01$, and decreases with the increasing $T_{\rm s}$.} 
	\label{fig: wcfit}
\end{figure*}

\begin{table*}
        \centering
	\caption{Reasonable vertical Thomson optical depths ($\tau_{\rm r}$) in simulations. The ``reasonable'' here means that when the fitting method works well, the best-fitting parameters of the output spectra should be normal in observations.}
	\label{tab:rea_ran}
	\begin{tabular}{cccccc}
		\hline
		~ & $kT_{\rm e}$/keV & 0.1 & 0.2 & 0.3 & 0.4  \\
		\hline
		~ & $\tau_{\rm r}$ & [30, 65] & [25, 45] & [20, 35] & [20, 30] \\	
		
		\hline
	\end{tabular}
\end{table*}

For the warm corona model, the best-fitting spectral parameters of the numerical simulation results are shown in Fig.~\ref{fig: wcfit}. The fitting results for $\tau=5$ are not shown here due to too less photon count in the fitting bands. Our simulation range of optical depth is much larger than the reasonable ranges constrained by observations, and therefore almost all the curves of best-fitting parameters more or less locate at the fitting boundary. From the $kT_{\rm d}=0.01$ keV (solid) fittings in the $kT_{\rm s}=0.01$ keV (left column) cases, one can identify the reasonable range, where $\Gamma$ varies with $\tau$ and $T_{\rm wc}$ roughly matches the $T_{\rm e}$. The reasonable ranges of $\tau$ for different $T_{\rm e}$ are listed in Table \ref{tab:rea_ran}, noticed that these values are just some rough estimates.

In the left column of Fig.~\ref{fig: wcfit}, $kT_{\rm s}=0.01$ keV, when $kT_{\rm d}$ increases from 0.01 (solid) to 0.03 keV (dashed), the fitting parameters are still stable, but when $kT_{\rm d}$ increases to 0.07 keV (dotted), the best-fitting parameters are counterfactual and oscillating (except the $kT_{\rm e}=0.01$ keV case, of which $T_{\rm wc}$ will naturally locate at the fitting boundary). In the middle column, $kT_{\rm s}=0.03$ keV, the overall curve profiles have not changed. However, the $T_{\rm wc}$ of the $kT_{\rm d}=0.01,~0.03$ (solid, dashed) fittings in the reasonable ranges are obviously lower than $T_{\rm e}$ when the corresponding $\Gamma$ do not reach the fitting boundary, especially for the $kT_{\rm e}=0.3,~0.4$ keV cases. Meanwhile, the $\Gamma$ also decrease. In the right column, $kT_{\rm s}=0.07$ keV, the best-fitting parameters are quite different. Now the fitting method hardly works when $kT_{\rm d}=0.01,~0.03$ (solid, dashed), one phenomenon is that their $\Gamma$ are almost 2.0 (lower boundary) everywhere. Meanwhile, for the $kT_{\rm d}=0.07$ (dotted) fittings, the parameters seem to be normal, just like the $kT_{\rm d}=0.03$ (dashed) fittings in the middle column, but indeed the $T_{\rm wc}$ in the reasonable ranges are also obviously underestimated. 

At the end of the discussion for Fig.~\ref{fig: wcfit}, we should mention that the fitting band (in different colours) does not affect the above conclusions: in most of the cases, the differences caused by the fitting band are significant only when at least one of the $T_{\rm wc},~\Gamma$ reaches the fitting boundaries, i.e., when the fitting method works poorly; the only exceptions come from the $kT_{\rm d}=0.07$ keV fittings in the $kT_{\rm s}=0.01,~0.03$ keV cases, as the red dotted curves show some differences, but in actual fittings $T_{\rm s}$ will be only underestimated. 

\begin{figure*}[b]
	\centering
	\includegraphics[width=1.0\textwidth]{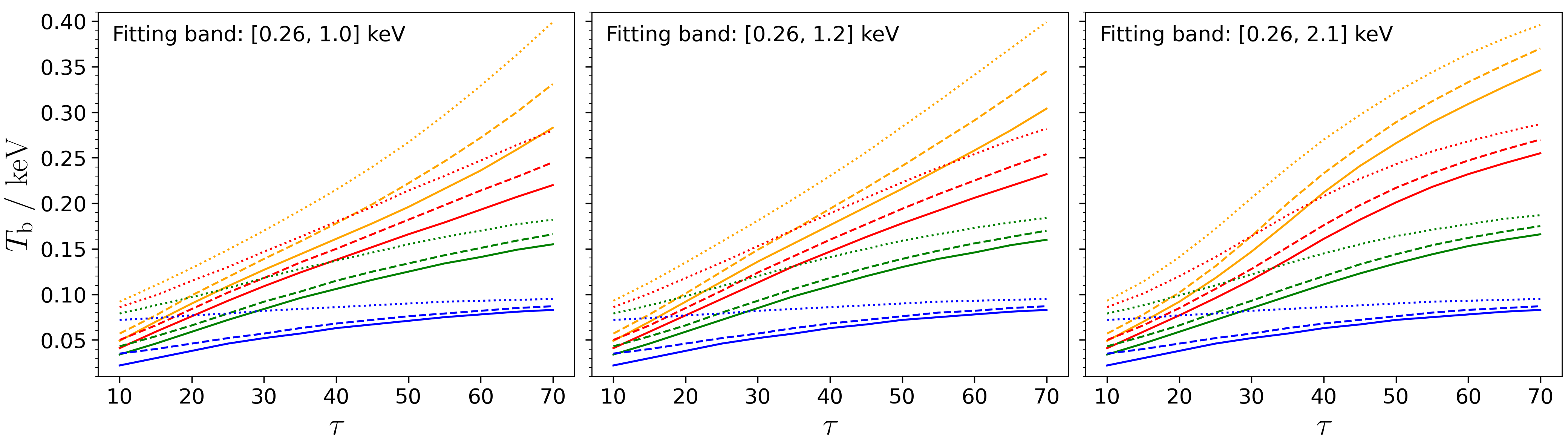}
	
	\caption{Best-fitting blackbody temperatures ($kT_{\rm b}$) of the numerical simulation results. $\tau$ is the vertical Thomson optical depth. The simulation parameters are: $kT_{\rm s}=$0.01 (solid), 0.03 (dashed), 0.07 (dotted) keV, and $kT_{\rm e}=$0.1 (blue), 0.2 (green), 0.3 (red), 0.4 (orange) keV. The fitting bands are listed in the corresponding panels. In general, $T_{\rm b}$ increases with $T_{\rm s}$ and $T_{\rm e}$, but can hardly exceed the corresponding $T_{\rm e}$. The fitting results are insensitive to the fitting band.} 
	\label{fig: bbfit}
\end{figure*}

For the blackbody model, the best-fitting blackbody temperatures ($kT_{\rm b}$) of the numerical simulation results are shown in Fig.~\ref{fig: bbfit}. In general, $T_{\rm b}$ increases with $T_{\rm s}$ and $T_{\rm e}$, but can hardly exceed the corresponding $T_{\rm e}$: for $kT_{\rm e}\leq0.2$ keV cases, the upper limitations of $T_{\rm b}$ seem to be lower than the corresponding $T_{\rm e}$; for $T_{\rm e}\geq 0.2$ keV cases, anyway $T_{\rm b}$ do not exceed $T_{\rm e}$ in simulation ranges. In the reasonable ranges of optical depth, $T_{\rm b}$ in all the cases can vary from a few tens of eVs to about 0.2 keV. It is worth noting that, unlike the warm corona model, the fitting results of the blackbody model are insensitive to the fitting band.

\begin{figure*}
	\centering
	\includegraphics[width=1.0\textwidth]{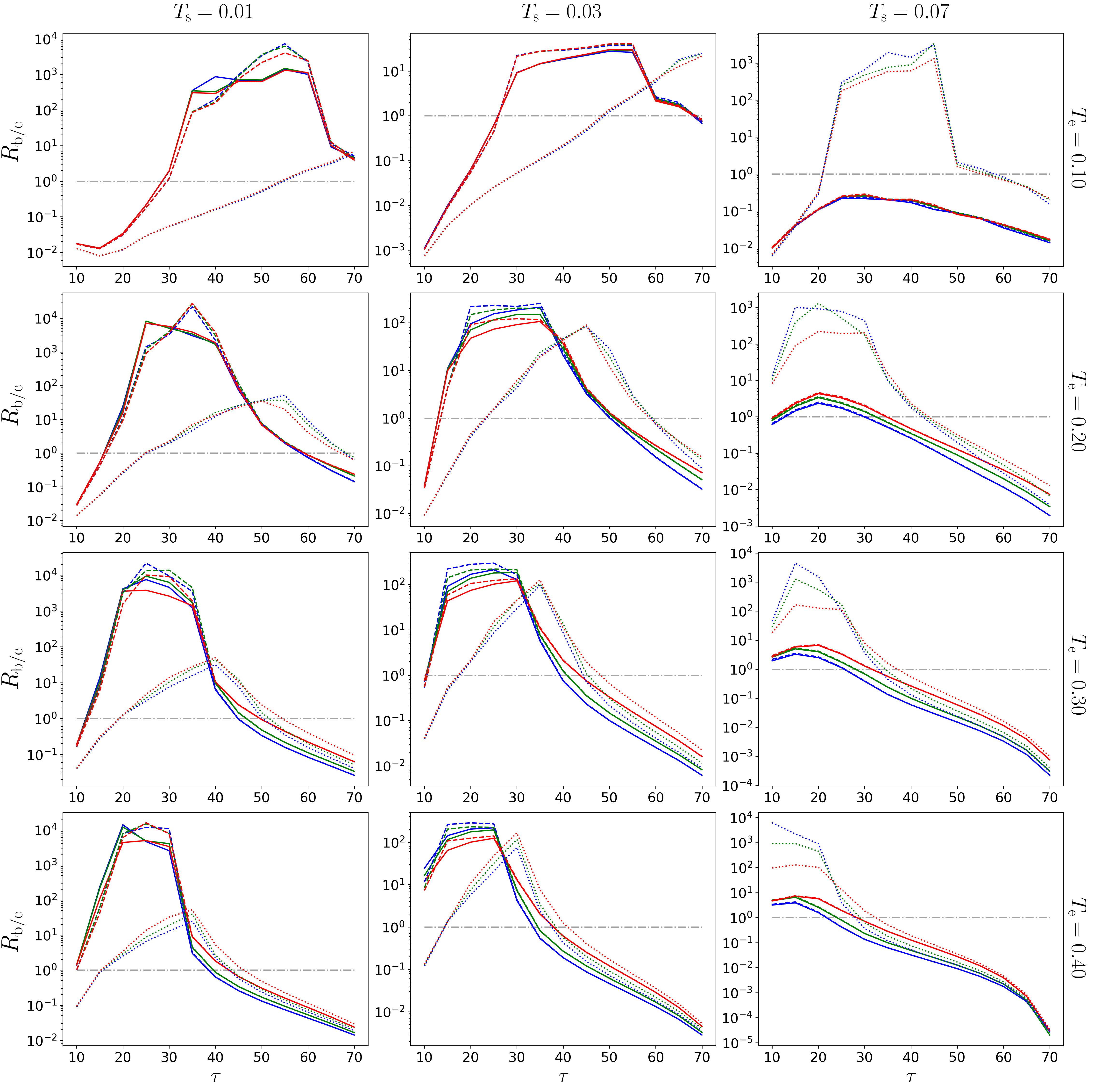}
	\caption{The ratio of $\chi^2$ ($R_{\rm b/c}$) between the warm corona model and blackbody model. $R_{\rm b/c}>1$ indicates that the warm corona model fits better. $\tau$ is the vertical Thomson optical depth. The simulation parameters are $kT_{\rm s}=0.01,~0.03,~0.07$ keV (from left to right cases), and $kT_{\rm e}=0.1,~0.2,~0.3,~0.4$ keV (from top to bottom cases). The fitting disc temperature $kT_{\rm d}=$ 0.01 (solid), 0.03 (dashed), 0.07 (dotted) keV, and the upper boundaries of the fitting band are 1.0 (blue), 1.2 (green), 2.1 (red) keV. The horizontal grey dashed-dotted lines correspond to $R_{\rm b/c}=1$. The results indicate that considerable spectra are blackbody-like when $kT_{\rm s}=0.07$ keV.} 
	\label{fig: x2r}
\end{figure*}

It is important to test which model fits the results better, and the results are shown in Fig.~\ref{fig: x2r}. $R_{\rm b/c}>1$ indicates that the warm corona model fits better, while $R_{\rm b/c}<1$ phenomenologically prefers the blackbody model. Here we should mention that, the behavior of the warm corona model will be better if we expand the fitting range of $\Gamma$, but the unusual parameters may not be very convincible in actual observational works. 

As shown in Fig.~\ref{fig: x2r}, when both $kT_{\rm s}$ and $kT_{\rm d}$ are $\leq0.03$ keV (solid, dashed curves in the left, middle columns), each fitting curve has a top platform almost overlapping the reasonable range, and $R_{\rm b/c}$ is still larger than 1 when it drops at the edges of the reasonable range. The character suggests that when $kT_{\rm s}$ is at most moderately high, the warm corona model is at least better within the reasonable range. However, it is clear that the $R_{\rm b/c}$ of the platform decreases with the increasing $kT_{\rm s}$, and the top platforms move to small $\tau$. As a result, when $kT_{\rm s}=kT_{\rm d}=0.07$ keV (dotted curves in the right column), we find that the blackbody model fits better at large reasonable $\tau$. Moreover, if $kT_{\rm d}<kT_{\rm s}=0.07$ keV (solid, dashed curves in the right column), the blackbody model will be better in a wide range, especially for the low $T_{\rm e}$ cases.

\section{Discussions}
\label{sec:discuss}

\subsection{Decrease of the best-fitting $T_{\rm wc}$ and $\Gamma$}
Based on the results shown in Fig. \ref{fig: wcfit}, it is clear that for electron temperatures ($kT_{\rm e}$) greater than 0.01 keV, the \textsc{thcomp} model fails to fit the spectra well, regardless of whether the seed photon temperature ($T_{\rm s}$) is accurately estimated or not. This is evident from the decreasing values of the best-fitting parameters $T_{\rm wc}$ and $\Gamma$ with increasing $T_{\rm s}$. The significant decrease suggests that the effect is primarily due to the seed photon temperature rather than variations in the details of simulations or fittings. Moreover, given the generality of these conclusions (different $T_{\rm e}$ and $T_{\rm s}$), a similar decrease in the fitting parameters can be expected for warm coronae with vertical structures and/or radius-dependent seed photon temperature.

One interesting thing is that the underestimated $kT_{\rm wc}$ are around 0.2 keV, which has been suggested as the most common warm coronal temperature \citep{Petrucci2018,Kubota2018,Mitchell2023}. Noticed that a low temperature warm corona must have a quite large optical depth to produce a flat spectrum (see Fig.~\ref{fig: wcfit}). Therefore, if the underestimation is widespread in observational works, then the required maximum optical depth will decrease. The decrease is favoured by simulation works, because a stable Comptonisation-dominant region on the disc surface can not be too thick (conservatively, can hardly reach $\tau=30$) if it keeps quasi-uniform temperature \citep{Rozanska2015,Ballantyne2020}, or the warm corona is thick and thermal unstable \citep{Gronkiewicz2023}, but now the vertical temperature variation is too drastic, and therefore the apparent temperature deviating further from the average temperature, the radiation spectrum does not match the profile of SE (Tang et al., {\it in prep.}).

A temperature of 0.03 keV is indeed a very common and typical innermost disc temperature for AGNs, but of course not all the discs can touch it. Therefore, the associated phenomena should have an impact on the fitting results of different samples. We notice that \cite{Petrucci2018} provided fitting results of a sample with simultaneous and both high-quality data in the optical/UV and X-ray bands observed by {\it XMM-Newton} \citep{Struder2001,Mason2001}. Their fitting model assumes that the coronae cover the entire disc, resulting in an inevitable lower fitting temperature for the seed photons. This could potentially underestimate the actual temperature. Their median best-fitting $kT_{\rm wc}$ is 0.24 keV, and the median best-fitting $\Gamma$ is 2.61 (Appendix C in the original paper). As a comparison, recently \cite{Waddell2023} provided a spectral fitting work for a hard X-ray–selected sample from the eFEDS (eROSITA Final Equatorial Depth Survey) sample \citep{Merloni2012,Predehl2021,Sunyaev2021,Brunner2022,Liu2022,Salvato2022}. AGNs in the hard X-ray-selected sample have a high X-ray flux \citep{Nandra2024}. Based on this, we consider AGNs in this sample to have a stronger coronal emission and a weaker disc emission. Consequently, we expect a lower actual $T_{\rm s}$ and an increase in the best-fitting parameters. For the warm coronal fittings (Table B.5 in the original paper), the median best-fitting $kT_{\rm wc}$ is 0.47 keV, and the median best-fitting $\Gamma$ is 3.15. Such high values agree with our prediction.

\subsection{Blackbody-like SE}
From Fig.~\ref{fig: x2r}, it can be observed that AGNs with extremely high seed photon temperatures ($T_{\rm s}$) have the potential to produce blackbody-like SEs. 
However, whether it is the analysis of individual sources \citep[e.g.][]{Xu2021} or statistical work on large samples \citep[e.g.][]{Waddell2023}, there is no evidence to suggest that a blackbody is a better fitting model than the warm corona. Then, since the physical warm corona models suggest that the warm coronal radius decreases with the increasing Eddington luminosity \citep{Kubota2018, Kubota2019}, we further consider Narrow-line Seyfert 1 galaxies (NLS1s), a subclass of AGNs believed to host relatively low-mass supermassive black holes and have high luminosities \citep{Osterbrock1985, Gu2015}. Based on the standard disc model, NLS1s are expected to have relatively higher seed photon temperatures within their compact coronal region, and therefore have blackbody-like or at least flat SEs. However, observations do not agree with this, the soft photon indices of NLS1s are not flatter than those of Broad-line Seyfert 1 galaxies \citep{Boller1996,Matzeu2020,Middei2020,Yu2023}.

The absence of a blackbody-like SE indicates that the temperature of seed photons may not be able to touch the extremely high temperature allowed by the standard disc model. One possible explanation is that energy is carried away by outflows/winds. For example, NLS1s are known to exhibit strong outflows. Moreover, \cite{Cai2023} recently reported that the average spectral energy distribution for quasars is much redder than prediction of the standard disc model, suggesting prevalent winds in quasars. Indeed, according to theoretical understanding, it is known that radiation-driven outflows are inevitably generated in high-temperature discs \citep[e.g.][]{Feng2019}. Meanwhile, the relation between warm coronae and outflows has also been found in simulations \citep[e.g.][]{Rozanska2015}.  

Based on the above discussion, we suggest that a physical model should incorporate an upper limit for the seed photon temperature (but the specific value still requires further investigation). This upper limit will modify the distribution of coronal radiation energy between the EUV and soft X-ray bands, and further affect the model-dependent warm coronal radius and intrinsic luminosity of AGNs.

\section{Conclusions}
\label{sec:conclu}
In this work, we study how the high seed photon temperature affect the soft excess in AGNs. We use the Monte Carlo method to simulate the Comptonisation process within the warm corona in parameter ranges of $kT_{\rm e}\in[0.1, 0.4]~{\rm keV}$, $\tau\in[5, 70]$, and $kT_{\rm s}=0.01,~0.03,~0.07$ keV. We then fit the output spectra by \textsc{thcomp} in parameter ranges of $kT_{\rm wc}\in[0.1, 1]~{\rm keV}$, $\Gamma\in[2, 3.5]$, and $kT_{\rm d}=0.01,~0.03,~0.07$ keV. We also fit the output spectra by a thermal component. There are two main conclusions:
\begin{itemize}	
	\item When $kT_{\rm s}$ is moderately high (0.03 keV in our simulations), the best-fitting $T_{\rm wc}$ and $\Gamma$ decrease with the increasing $T_{\rm s}$. The apparent low-temperature and flat SE probably leads to mistaking 0.2 keV as the most common $T_{\rm e}$ and overestimating the optical thickness of warm coronae. Our results also indicate that low $T_{\rm s}$ will be more likely to lead to high $T_{\rm wc}$ and steep $\Gamma$. This phenomenon can be used to explain why the high X-ray flux AGN sample \citep{Waddell2023} exhibits higher best-fitting $kT_{\rm wc}$ and $\Gamma$ compared to the multi-band-selected AGN sample \citep{Petrucci2018}.
	\item When $kT_{\rm s}$ is extremely high, i.e., very close to 0.1 keV (0.07 keV in our simulations), the SE profile will be more likely to be relatively blackbody-like. The absence of a blackbody-like SE in actual AGN spectra indicates that the temperature of seed photons is not able to touch the extremely high temperature allowed by the standard disc model. Therefore, a physical warm corona model should incorporate an upper limit for the seed photon temperature.
\end{itemize}

Our conclusions indicate that, for spectral fittings of the warm coronal radiation (SE in AGNs), $kT_{\rm s}$ should be treated as a free parameter with an upper limit, and an accurate coronal geometry is necessary when $kT_{\rm s}>0.01$ keV.

\normalem
\begin{acknowledgements}
We thank Mr. Scott Hagen (University of Durham) for advise on warm corona model. We thanks Dr. Jun Li (Guangzhou University) for help on numerical simulation. The work is partially supported by the National Natural Science Foundation of China (NSFC U2031201 and NSFC 11733001) and the Natural Science Foundation of Guangdong Province (2019B030302001). We acknowledge the science research grants from the China Manned Space Project, with NO. CMS-CSST-2021-A06. We also acknowledge support from Astrophysics Key Subjects of Guangdong Province and Guangzhou City, and support from Scientific and Technological Cooperation Projects (2020-2023) between the People's Republic of China and the Republic of Bulgaria, as well as support that was received from Guangzhou University (No. YM2020001).

\end{acknowledgements}
  
\bibliographystyle{raa}
\bibliography{ref}

\begin{thebibliography}{64}
\providecommand\natexlab[1]{#1}
\providecommand\JournalTitle[1]{#1}

\bibitem[{Arnaud}(1996)]{XSPEC}
{Arnaud}, K.~A. 1996, in Astronomical Society of the Pacific Conference Series,
  Vol. 101, Astronomical Data Analysis Software and Systems V, ed. G.~H.
  {Jacoby} \& J.~{Barnes}, 17

\bibitem[{Ballantyne}(2020)]{Ballantyne2020}
{Ballantyne}, D.~R. 2020, \mnras, 491, 3553

\bibitem[{Ballantyne} \& {Xiang}(2020)]{Ballantyne2020b}
{Ballantyne}, D.~R., \& {Xiang}, X. 2020, \mnras, 496, 4255

\bibitem[{Boissay} {et~al.}(2016)]{Boissay2016}
{Boissay}, R., {Ricci}, C., \& {Paltani}, S. 2016, \aap, 588, A70

\bibitem[{Boller} {et~al.}(1996)]{Boller1996}
{Boller}, T., {Brandt}, W.~N., \& {Fink}, H. 1996, \aap, 305, 53

\bibitem[{Brunner} {et~al.}(2022)]{Brunner2022}
{Brunner}, H., {Liu}, T., {Lamer}, G., {et~al.} 2022, \aap, 661, A1

\bibitem[{Cai} \& {Wang}(2023)]{Cai2023}
{Cai}, Z.-Y., \& {Wang}, J.-X. 2023, Nature Astronomy, 7, 1506

\bibitem[{Crummy} {et~al.}(2006)]{Crummy2006}
{Crummy}, J., {Fabian}, A.~C., {Gallo}, L., \& {Ross}, R.~R. 2006, \mnras, 365,
  1067

\bibitem[{De Marco} {et~al.}(2013)]{Marco2013}
{De Marco}, B., {Ponti}, G., {Cappi}, M., {et~al.} 2013, \mnras, 431, 2441

\bibitem[{Dolence} {et~al.}(2009)]{Dolence2009}
{Dolence}, J.~C., {Gammie}, C.~F., {Mo{\'s}cibrodzka}, M., \& {Leung}, P.~K.
  2009, \apjs, 184, 387

\bibitem[{Done} {et~al.}(2012)]{Done2012}
{Done}, C., {Davis}, S.~W., {Jin}, C., {Blaes}, O., \& {Ward}, M. 2012, \mnras,
  420, 1848

\bibitem[{Feng} {et~al.}(2019)]{Feng2019}
{Feng}, J., {Cao}, X., {Gu}, W.-M., \& {Ma}, R.-Y. 2019, \apj, 885, 93

\bibitem[{Garc{\'\i}a} {et~al.}(2019)]{Garcia2019}
{Garc{\'\i}a}, J.~A., {Kara}, E., {Walton}, D., {et~al.} 2019, \apj, 871, 88

\bibitem[{Gronkiewicz} {et~al.}(2023)]{Gronkiewicz2023}
{Gronkiewicz}, D., {R{\'o}{\.z}a{\'n}ska}, A., {Petrucci}, P.-O., \& {Belmont},
  R. 2023, \aap, 675, A198

\bibitem[{Gu} {et~al.}(2015)]{Gu2015}
{Gu}, M., {Chen}, Y., {Komossa}, S., {et~al.} 2015, \apjs, 221, 3

\bibitem[{Hagen} \& {Done}(2023)]{Hagen2023}
{Hagen}, S., \& {Done}, C. 2023, \mnras, 525, 3455

\bibitem[{Jiang} {et~al.}(2018)]{Jiang2018}
{Jiang}, J., {Parker}, M.~L., {Fabian}, A.~C., {et~al.} 2018, \mnras, 477, 3711

\bibitem[{Jin} {et~al.}(2022)]{Jin2022}
{Jin}, C., {Done}, C., {Ward}, M., {et~al.} 2022, \mnras, 512, 5642

\bibitem[{Jin} {et~al.}(2012)]{Jin2012}
{Jin}, C., {Ward}, M., {Done}, C., \& {Gelbord}, J. 2012, \mnras, 420, 1825

\bibitem[{Kara} {et~al.}(2016)]{Kara2016}
{Kara}, E., {Alston}, W.~N., {Fabian}, A.~C., {et~al.} 2016, \mnras, 462, 511

\bibitem[{Kompaneets}(1957)]{Kompaneets1957}
{Kompaneets}, A.~S. 1957, Soviet Journal of Experimental and Theoretical
  Physics, 4, 730

\bibitem[{Kubota} \& {Done}(2018)]{Kubota2018}
{Kubota}, A., \& {Done}, C. 2018, \mnras, 480, 1247

\bibitem[{Kubota} \& {Done}(2019)]{Kubota2019}
{Kubota}, A., \& {Done}, C. 2019, \mnras, 489, 524

\bibitem[{Laurenti} {et~al.}(2022)]{Laurenti2022}
{Laurenti}, M., {Piconcelli}, E., {Zappacosta}, L., {et~al.} 2022, \aap, 657,
  A57

\bibitem[{Li} {et~al.}(2010)]{li2010}
{Li}, G.-X., {Yuan}, Y.-F., \& {Cao}, X. 2010, \apj, 715, 623

\bibitem[{Liu} {et~al.}(2017)]{Liu2017}
{Liu}, T., {Tozzi}, P., {Wang}, J.-X., {et~al.} 2017, \apjs, 232, 8

\bibitem[{Liu} {et~al.}(2022)]{Liu2022}
{Liu}, T., {Buchner}, J., {Nandra}, K., {et~al.} 2022, \aap, 661, A5

\bibitem[{Mallick} {et~al.}(2021)]{Mallick2021}
{Mallick}, L., {Wilkins}, D.~R., {Alston}, W.~N., {et~al.} 2021, \mnras, 503,
  3775

\bibitem[{Mason} {et~al.}(2001)]{Mason2001}
{Mason}, K.~O., {Breeveld}, A., {Much}, R., {et~al.} 2001, \aap, 365, L36

\bibitem[{Matt} {et~al.}(2014)]{Matt2014}
{Matt}, G., {Marinucci}, A., {Guainazzi}, M., {et~al.} 2014, \mnras, 439, 3016

\bibitem[{Matzeu} {et~al.}(2020)]{Matzeu2020}
{Matzeu}, G.~A., {Nardini}, E., {Parker}, M.~L., {et~al.} 2020, \mnras, 497,
  2352

\bibitem[{Mehdipour} {et~al.}(2011)]{Mehdipour2011}
{Mehdipour}, M., {Branduardi-Raymont}, G., {Kaastra}, J.~S., {et~al.} 2011,
  \aap, 534, A39

\bibitem[{Mehdipour} {et~al.}(2015)]{Mehdipour2015}
{Mehdipour}, M., {Kaastra}, J.~S., {Kriss}, G.~A., {et~al.} 2015, \aap, 575,
  A22

\bibitem[{Merloni} {et~al.}(2012)]{Merloni2012}
{Merloni}, A., {Predehl}, P., {Becker}, W., {et~al.} 2012, arXiv e-prints,
  arXiv:1209.3114

\bibitem[{Middei} {et~al.}(2020)]{Middei2020}
{Middei}, R., {Petrucci}, P.~O., {Bianchi}, S., {et~al.} 2020, \aap, 640, A99

\bibitem[{Mitchell} {et~al.}(2023)]{Mitchell2023}
{Mitchell}, J. A.~J., {Done}, C., {Ward}, M.~J., {et~al.} 2023, \mnras, 524,
  1796

\bibitem[Nandra {et~al.}(2024)]{Nandra2024}
Nandra, K., Waddell, S. G.~H., Liu, T., {et~al.} 2024, The eROSITA Final
  Equatorial Depth Survey (eFEDS): the hard X-ray selected sample,
  arXiv:2401.17300

\bibitem[{Noda} {et~al.}(2013)]{Noda2013}
{Noda}, H., {Makishima}, K., {Nakazawa}, K., {et~al.} 2013, \pasj, 65, 4

\bibitem[{Novikov} \& {Thorne}(1973)]{Novikov1973}
{Novikov}, I.~D., \& {Thorne}, K.~S. 1973, in Black Holes (Les Astres Occlus),
  343

\bibitem[{Osterbrock} \& {Pogge}(1985)]{Osterbrock1985}
{Osterbrock}, D.~E., \& {Pogge}, R.~W. 1985, \apj, 297, 166

\bibitem[{Petrucci} {et~al.}(2018)]{Petrucci2018}
{Petrucci}, P.~O., {Ursini}, F., {De Rosa}, A., {et~al.} 2018, \aap, 611, A59

\bibitem[{Petrucci} {et~al.}(2013)]{Petrucci2013}
{Petrucci}, P.~O., {Paltani}, S., {Malzac}, J., {et~al.} 2013, \aap, 549, A73

\bibitem[{Petrucci} {et~al.}(2020)]{Petrucci2020}
{Petrucci}, P.~O., {Gronkiewicz}, D., {Rozanska}, A., {et~al.} 2020, \aap, 634,
  A85

\bibitem[{Porquet} {et~al.}(2018)]{Porquet2018}
{Porquet}, D., {Reeves}, J.~N., {Matt}, G., {et~al.} 2018, \aap, 609, A42

\bibitem[{Predehl} {et~al.}(2021)]{Predehl2021}
{Predehl}, P., {Andritschke}, R., {Arefiev}, V., {et~al.} 2021, \aap, 647, A1

\bibitem[Reynolds(2021)]{Reynolds2020}
Reynolds, C.~S. 2021, Ann. Rev. Astron. Astrophys., 59, 117

\bibitem[{Ross} \& {Fabian}(2005)]{Ross2005}
{Ross}, R.~R., \& {Fabian}, A.~C. 2005, \mnras, 358, 211

\bibitem[{R{\'o}{\.z}a{\'n}ska} {et~al.}(2015)]{Rozanska2015}
{R{\'o}{\.z}a{\'n}ska}, A., {Malzac}, J., {Belmont}, R., {Czerny}, B., \&
  {Petrucci}, P.~O. 2015, \aap, 580, A77

\bibitem[{Salvato} {et~al.}(2022)]{Salvato2022}
{Salvato}, M., {Wolf}, J., {Dwelly}, T., {et~al.} 2022, \aap, 661, A3

\bibitem[{Shakura} \& {Sunyaev}(1973)]{Shakura1973}
{Shakura}, N.~I., \& {Sunyaev}, R.~A. 1973, \aap, 500, 33

\bibitem[{Str{\"u}der} {et~al.}(2001)]{Struder2001}
{Str{\"u}der}, L., {Briel}, U., {Dennerl}, K., {et~al.} 2001, \aap, 365, L18

\bibitem[{Sunyaev} \& {Titarchuk}(1980)]{Sunyaev1980}
{Sunyaev}, R.~A., \& {Titarchuk}, L.~G. 1980, \aap, 86, 121

\bibitem[{Sunyaev} {et~al.}(2021)]{Sunyaev2021}
{Sunyaev}, R., {Arefiev}, V., {Babyshkin}, V., {et~al.} 2021, \aap, 656, A132

\bibitem[{Tang} {et~al.}(2023)]{Tang2023}
{Tang}, Z.-Y., {Feng}, J.-J., \& {Fan}, J.-H. 2023, \mnras, 520, 129

\bibitem[{Waddell} {et~al.}(2023)]{Waddell2023}
{Waddell}, S. G.~H., {Nandra}, K., {Buchner}, J., {et~al.} 2023, arXiv
  e-prints, arXiv:2306.00961

\bibitem[{Walter} \& {Fink}(1993)]{Walter1993}
{Walter}, R., \& {Fink}, H.~H. 1993, \aap, 274, 105

\bibitem[{Wu} {et~al.}(2015)]{Wu2015}
{Wu}, X.-B., {Wang}, F., {Fan}, X., {et~al.} 2015, \nat, 518, 512

\bibitem[{Xu} {et~al.}(2021{\natexlab{a}})]{Xu2021}
{Xu}, X., {Ding}, N., {Gu}, Q., {Guo}, X., \& {Contini}, E. 2021{\natexlab{a}},
  \mnras, 507, 3572

\bibitem[{Xu} {et~al.}(2021{\natexlab{b}})]{XuYR2021}
{Xu}, Y., {Garc{\'\i}a}, J.~A., {Walton}, D.~J., {et~al.} 2021{\natexlab{b}},
  \apj, 913, 13

\bibitem[{Yu} {et~al.}(2023)]{Yu2023}
{Yu}, Z., {Jiang}, J., {Bambi}, C., {et~al.} 2023, \mnras, 522, 5456

\bibitem[{Zdziarski} {et~al.}(1996)]{Zdziarski1996}
{Zdziarski}, A.~A., {Johnson}, W.~N., \& {Magdziarz}, P. 1996, \mnras, 283, 193

\bibitem[{Zdziarski} {et~al.}(2020)]{Zdziarski2020}
{Zdziarski}, A.~A., {Szanecki}, M., {Poutanen}, J., {Gierli{\'n}ski}, M., \&
  {Biernacki}, P. 2020, \mnras, 492, 5234

\bibitem[{Zoghbi} \& {Miller}(2023)]{Zoghbi2023}
{Zoghbi}, A., \& {Miller}, J.~M. 2023, \apj, 957, 69

\bibitem[{{\.Z}ycki} {et~al.}(1999)]{Zycki1999}
{{\.Z}ycki}, P.~T., {Done}, C., \& {Smith}, D.~A. 1999, \mnras, 309, 561

\end{thebibliography}

\end{document}